# A Mapping Sheath with Thermally Drawn Multi-Electrode Basket for Cardiac Electrophysiological Recording and Ablation Catheter Delivery

Qindong Zheng[1], Anil Demircali[2], Jinshi Zhao[2], Xiaotong Guo[3], Libaihe Tian[2], Oliver Jones[1,4], Jamie Kay[1,4], Shengzhe Li[5], Alex Ranne[6], Elaine Lim[4], Huiyi Wu[4], Simos Koutsoftidis[1], Mohamed Abdelaziz[4], Emmanuel Drakakis[1], Prapa Kanagaratnam[4], Nick Linton[1,4*], Burak Temelkuran[2*]

[1]Department of Bioengineering, Imperial College London, London, UK
[2]Department of Metabolism, Digestion and Reproduction, Imperial College London, London, UK
[3]Department of Electrical and Electronic Engineering, Imperial College London, London, UK
[4]National Heart & Lung Institute, London, UK
[5]Haihe Laboratory of Brain-Computer Interaction and Human-Machine Integration, Tianjin, China
[6]Department of Mechanical Engineering, Imperial College London, London, UK
[*] Corresponding authors. Email: nick.linton@imperial.ac.uk; b.temelkuran@imperial.ac.uk

## Abstract

Cardiac arrhythmias, particularly atrial fibrillation (AF), represent a major cardiovascular health burden and underscore the need for efficient and integrated strategies for electrical mapping and targeted therapy. Cardiac electrophysiology (EP) procedures depend on accurate identification of arrhythmogenic substrates followed by timely catheter ablation, but conventional diagnostic and therapeutic devices remain separate, often requiring repeated catheter exchanges and multiple access routes. Here, we report an adaptable strategy for functionalizing hollow-core sheaths with EP mapping capabilities, integrating multielectrode recording and ablation catheter delivery within a single compact platform. The device leverages thermal drawing to enable complex geometric fabrication, miniaturization, rapid prototyping, and scalable manufacturing of ultrathin electrode splines arranged circumferentially at the distal end to form an adjustable basket. The mapping sheath exhibited mechanical and electrophysiological properties suitable for intracardiac navigation and electrogram recording in bench-top evaluations, an in vitro left atrial phantom study, and ex vivo Langendorff-perfused porcine heart testing. In vivo porcine studies further demonstrated translational feasibility through vascular introduction, fluoroscopic visualization, intracardiac deployment, tissue contact, electrogram acquisition, and reconstruction of voltage and activation maps. These results support the development of intracardiac platforms with an adapted manufacturing approach, potentially guiding advances in agile cardiac mapping and ablation.

## 1. Introduction

Cardiac arrhythmias are a spectrum of abnormalities in cardiac rhythm, reflecting disruptions in the electrical conduction across the heart. Such perturbations in electrical conduction can compromise the physiological contractile function of the heart, manifesting clinically as a range of symptoms, including palpitations, dizziness, dyspnea, syncope, chest pain, and, in severe cases, sudden cardiac arrest (*1*). The most common arrhythmia is atrial fibrillation (AF), occurring in around 8% of people at the age of 80, and associated with a reduction in quality of life, stroke, and increased mortality (*2*). These arrhythmias are triggered by endocardial substrates that initiate and sustain abnormal electrical impulses within localized regions of cardiac tissue (*3*). Mapping endocardial electrical activity is necessary to determine the anatomic locations of these arrhythmia triggers, followed by terminating these triggers with the deployment of definitive ablation therapy (e.g. local cauterization of the tissue) (*4*). For patients with AF and left atrial regular tachycardias, the procedure requires dual access to the left side of the heart with both the mapping and ablation catheters passed separately through the inter-atrial septum, a thin muscular wall separating the left and right atria, requiring punctures during catheter access.

The approach of dual left atrial access for simultaneous mapping and ablation catheter deployment has served as the clinical "gold standard" for treating regular atrial tachycardia and atrial fibrillation driven by ectopic beats from pulmonary veins. Despite the reduction of procedure duration due to continuous potential monitoring, it demands extra time for creating two separate inter-atrial septum punctures compared with a single transseptal puncture (*5*). Additionally, cooperating with two independent catheters within the confined anatomical space of the left atrium frequently necessitates back-and-forth catheter withdrawals, introducing temporal latency during catheter exchanges. This temporal latency can restrict the timely delivery of ablation needed to terminate active arrhythmia triggers. These limitations become more pronounced in the context of complex arrhythmia management, such as persistent AF and scar-related tachycardia. These complex arrhythmias often arise from diverse intra-chamber sites and are characterized by highly heterogeneous activation patterns, dynamically shifting focal (localized rapid triggers) or re-entrance (self-sustaining loops), and marked beat-to-beat variability (*6*).

Another challenge for current arrhythmia treatment is the need to conform electrodes to various sites on endocardial substrates (*7*). Contemporary solutions involve instrumenting catheters with multiple electrodes arranged in various layouts, including loop, spiral, and grid configurations, each engineered to enhance tissue contact within distinct anatomical substrates, such as pulmonary veins and the posterior wall (*8*). Although these solutions enable simultaneous multi-electrode recording and mapping, they are limited by fixed geometric configurations that restrict their ability to conform dynamically during maneuvering and mapping across various anatomic geometries, which is essential for complex arrhythmia treatment (*9*). For instance, the demand for activation tracking from the posterior wall to the ostium and deep within the pulmonary vein during atrial fibrillation treatment necessitates the use of multiple mapping catheters in some clinical scenarios.

The basket mapping catheter with a deployable multi-spline structure has emerged as an alternative, offering adaptable electrode configurations that optimize endocardial contact (*10, 11*). The configurable deployment provided by basket catheters enables access to conventionally challenging regions, such as pulmonary veins, superior vena cava, and right ventricular outflow tract, while offering immediate assessment of ectopic-trigger isolation. Additionally, the broad spatial coverage of basket catheters enables tracking of complex arrhythmogenic drivers, including focal sources and rotors, in contrast to conventional localized catheters that detect such activity only when incidentally positioned over the arrhythmogenic core (*12*). Another approach that provides deployable configuration and global coverage is mapping with balloon catheters which form a high-resolution noncontact electrode array for deploying in all four cardiac chambers and reconstructing endocardial activation simultaneously (*13-17*). However, ballon catheters remain infrequently used in clinical practice, owing to progressive electrogram quality deterioration arising from increased electrode-tissue distance, limited conformity to the endocardium contour following balloon expansion, and potential risk of complications posed to patients (*18*).

An effective strategy to address these challenges is to integrate ablation and deployable mapping functions into one device, thereby maintaining intimate and adaptive contact with targeted endocardial regions while enabling timely ablation without an inter-atrial septum puncture and repeated catheter exchanges. Although several pulsed-field ablation (PFA) systems integrate mapping and ablation within a single catheter, clinical evidence remains centered on pulmonary vein isolation for AF (*19*). Evidence for other arrhythmias, including atrial and ventricular tachycardia, is comparatively limited. Moreover, the relative sparing of neural tissue by PFA may restrict its efficacy in complex arrhythmia involving autonomic substrates (*19*).

A central obstacle to advancing catheter technologies is the absence of rapid, cost-effective prototyping platforms capable of supporting iterative design and experimental validation. To address this gap, we introduce a customized catheter fabrication strategy based on a highly adaptable thermal drawing process that enables rapid, scalable, and cost-efficient prototyping of complex catheter architectures. In this process, a macroscopic preform, typically centimeters in diameter and engineered to replicate the target cross-sectional geometry, is thermally softened and drawn into a viscous state, yielding a continuous fiber that preserves the transverse architecture of the preform while achieving the requisite miniaturized dimensions (*20*). This adapted fabrication paradigm enables accelerated prototyping of complex catheter structures with fine control over geometry. The effectiveness of the thermal drawing technique in medical device development has been demonstrated across multiple applications, including cardiovascular interventions (*21*), medical robots (*22*), cochlear implants (*23*), and neurointerventional surgeries (*24*).

In this work, we present a novel vascular sheath with thermally drawn ultra-thin deployable electrode splines that seamlessly integrate electrophysiological mapping and ablation catheter delivery into a single, compact platform. The sheath features circumferentially deployable multi-electrode splines at its distal end and a central lumen sized to accommodate standard 8 Fr (2.67 mm) ablation catheters, while maintaining an overall profile below 14 Fr (4.67 mm) for compatibility with common introducers. The splines are fabricated using thermal drawing technique and integrated with bespoke thin electrodes. The distal segment of the device is engineered with enhanced flexibility and passively actuated by the inner steerable ablation catheter, offering high degrees of conformability and steerability. This structural and functional integration reduces the need for multiple device exchanges, streamlining interventional procedures. To evaluate device performance, we conducted *ex vivo* electrophysiological testing in a Langendorff-perfused porcine heart model, followed by *in vivo* validation in a large animal porcine model to further assess clinical feasibility. The results demonstrate the promise of this integrated platform as a next-generation tool for comprehensive cardiac mapping and targeted ablation therapy delivery.

## 2. Results

### 2.1. The Principle of Design

Inspired by the clinical challenges in arrhythmia treatment, our design features a deployable multi-electrode sheath that enables ablation catheter integration and agile electrophysiological mapping, hereafter referred to as a mapping sheath (Figure 1a). The primary objective of this design is to enable access to the cardiac chambers from the inferior vena cava through percutaneous femoral venous access. The distal end of the device must also be maneuverable into the left atrium by puncturing and crossing the interatrial septum, the thin muscular wall separating the left and right atria. This transseptal approach requires distal deflection exceeding 90° to facilitate septal puncture and left atrial entry. The deployable electrode architecture enables agile intracardiac mapping within the complex geometries of the cardiac chambers. In addition, the device should be compatible with commercially available ablation catheters, enabling intracardiac electrograms to be monitored during ablation therapy.

The mapping sheath is composed of a hollow-core shaft terminating in multiple flexible splines with electrode integrated, hereafter referred to as mapping splines, forming a deployable basket structure at its distal end. To align with catheter dimensions commonly used in clinical practice, the mapping sheath was designed with an overall profile compatible with standard 14 Fr introducers for cardiac access. Its hollow-core architecture enables the introduction of functional and deflectable catheters up to 8 Fr, including dilators, guidewires, transseptal needles, and ablation catheters (Figure 1b). The sheath body was engineered with dual stiffness by incorporating a variable-durometer braided tube as the structural backbone, providing distal segment with reduced stiffness for tip maneuverability driven by the deflectable catheter therein. As shown in Figure 1c, the proposed multi-spline basket structure enables circumferential electrode arrangement and deployment through relative axial sliding between the multilayer components of the sheath.

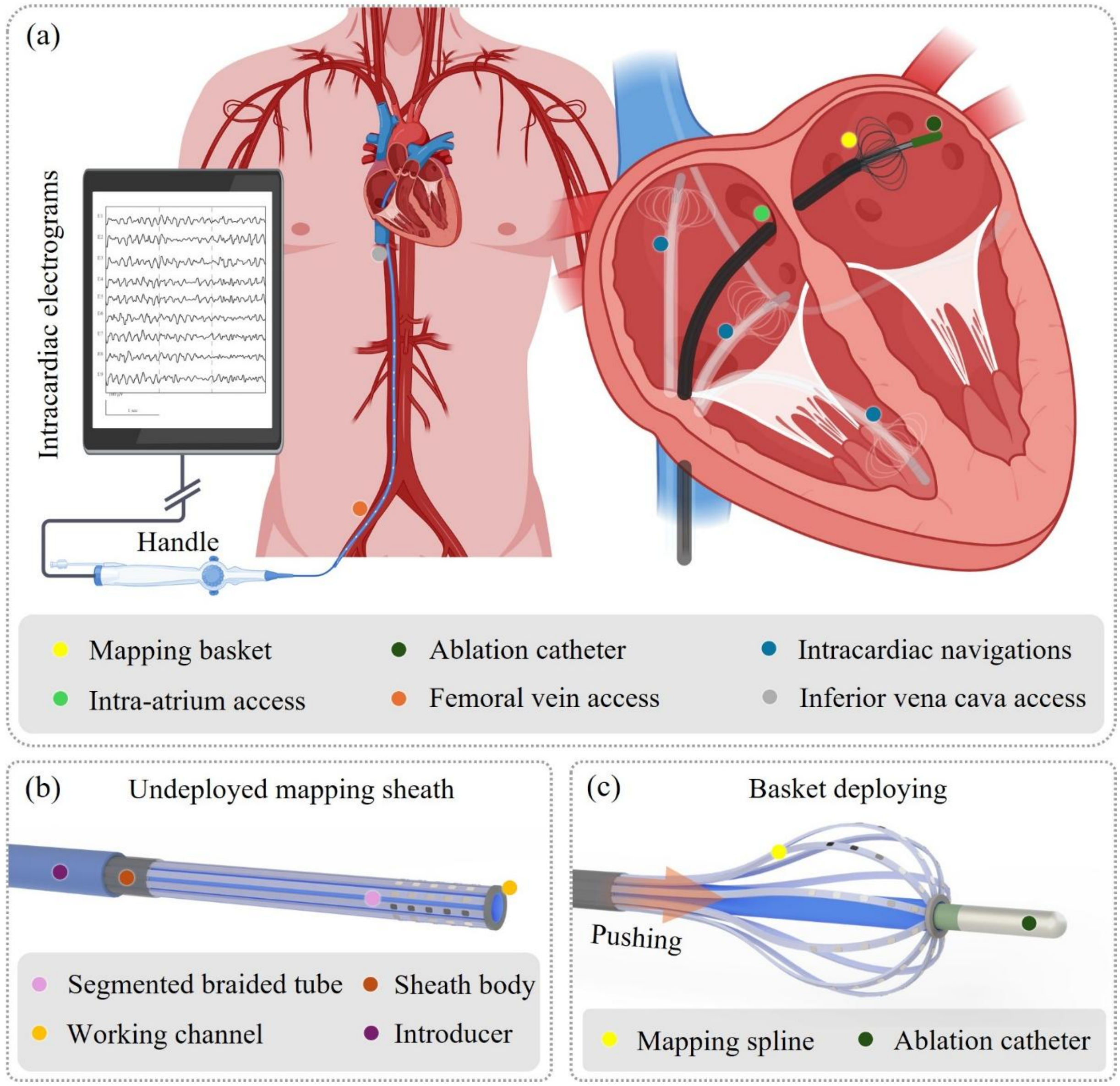

**Figure 1. Overview of the mapping sheath design.** (a) Schematic illustration of the clinical application of the proposed mapping catheter, enabling mapping and ablation. (b) Schematic demonstration of the mapping sheath distal tip under undeployed status. (c) Schematic demonstration of the mapping sheath distal tip under deployed status, forming a mapping basket.

## 2.2. Thermally Drawn Mapping Splines

The prototyping of mapping catheters has always pursued encapsulating and integrating multifunctional modules, including conductive wires, pulling tendons, and irrigation mechanisms, into a multi-lumen shaft fabricated with extrusion that is then bonded with a molded distal end (*25*). However, this approach is costly and not well-suited for rapid and iterative prototyping due to its demands for custom dies and tooling for each geometric modification. Additionally, the inherent limitation of extrusion further restrains the flexibility and miniaturization of designs with complex geometries. Thus, to fabricate the mapping splines with complex geometries and miniaturized dimensions, we introduced a rapid, scalable, and low-cost fabrication method for intraluminal devices based on a thermal drawing technique. Fabrication using this thermal drawing technique transforms macro-scale preform into a micro-scale while preserving complex cross-sectional features (Figure 2a). Precise control of the heating temperature and pulling speed enables scalable micron- to millimeter-scale fiber dimensions, a consistent cross-sectional profile, and a high manufacturing volume. Manufacturing volume can be scaled to kilometers per draw by using larger preforms for mass production. We further developed a wire-feeding platform to integrate wires directly into the fiber during thermal drawing.

To further improve prototyping iteration efficiency and customization flexibility, we incorporated high-resolution 3D printing as a rapid prototyping method for preform fabrication (Figure 2b), replacing the conventional molding and machining approaches. The preform material was selected as polycarbonate (PC) from various thermally drawable thermoplastic candidates (e.g., poly (methylmethacrylate) and acrylonitrile butadiene styrene) due to its biocompatibility, high mechanical strength, and dimensional stability. The PC preform is designed with a 40 mm x 12 mm rectangular structure, a 40-fold ratio to the spline's dimensions, and a length of 170 mm (Figure 2c). The resultant 3D printed preform possesses a tolerance within ±0.1mm in dimensions. A continuous fiber exceeding 30 meters in length was fabricated from a single preform, maintaining a tolerance in thickness within ±100 µm with the fiber drawing technique. A 1-meter segment with ±25 µm in thickness tolerance was then sectioned for the following assembly. Figure 2d demonstrates that the drawn fiber's cross-sectional geometry was well retained compared with the designated preform. The drawn fiber provides an overall cross-sectional dimension of 800 µm × 350 µm, five 120 µm square channels with conductive wires fed (enameled copper, 80 µm), and a 700 µm × 100 µm rectangle channel for nitinol backbone.

The mapping splines incorporated multiple miniaturized low-profile electrodes for electrogram acquisition while preserving a compact device profile. Platinum-iridium (PtIr) electrodes were selected due to their excellent biocompatibility, corrosion resistance, radiopacity, mechanical durability, and stable low-noise signal recording during electrophysiology (EP) mapping (*26*). The electrodes were fabricated by laser cutting a thin PtIr film into microscale features measuring 600 μm × 800 μm, with a thickness of 100 μm. Each electrode was laser-welded to an embedded conductive wire and positioned within pre-defined micromachined openings along the splines (Figure 2e). As shown in Figure 2f, the electrodes were arranged at close, uniform 2 mm intervals to enhance spatial resolution during electroanatomic mapping (*27*). A thin nitinol ribbon, 600 μm wide and 80 μm thick, was co-integrated within each spline to ensure stable deployment and consistent spatial positioning during cardiac navigation and mapping. Nitinol superelasticity provides sufficient structural rigidity for deployment while preserving flexibility within the dynamic cardiac environment (Figure 2g).

The mapping sheath assembly comprised a multilayer architecture that encapsulates the mapping splines, enables relative sliding during deployment, and preserves a compact form factor. As shown in Figures 2h and 2i, the assembled mapping splines were laminated between liner and heat-shrink layers, forming an individual layer. The braided tube served as the inner tube and structural backbone of the sheath body. A 50 mm segment at the distal end of the mapping sheath was designed as the basket-deployment section, where the exposed mapping splines could bend outward and inward during retracting and releasing the distal end with the liner. Detailed specifications of the sheath components are shown in Supplementary Table S1.

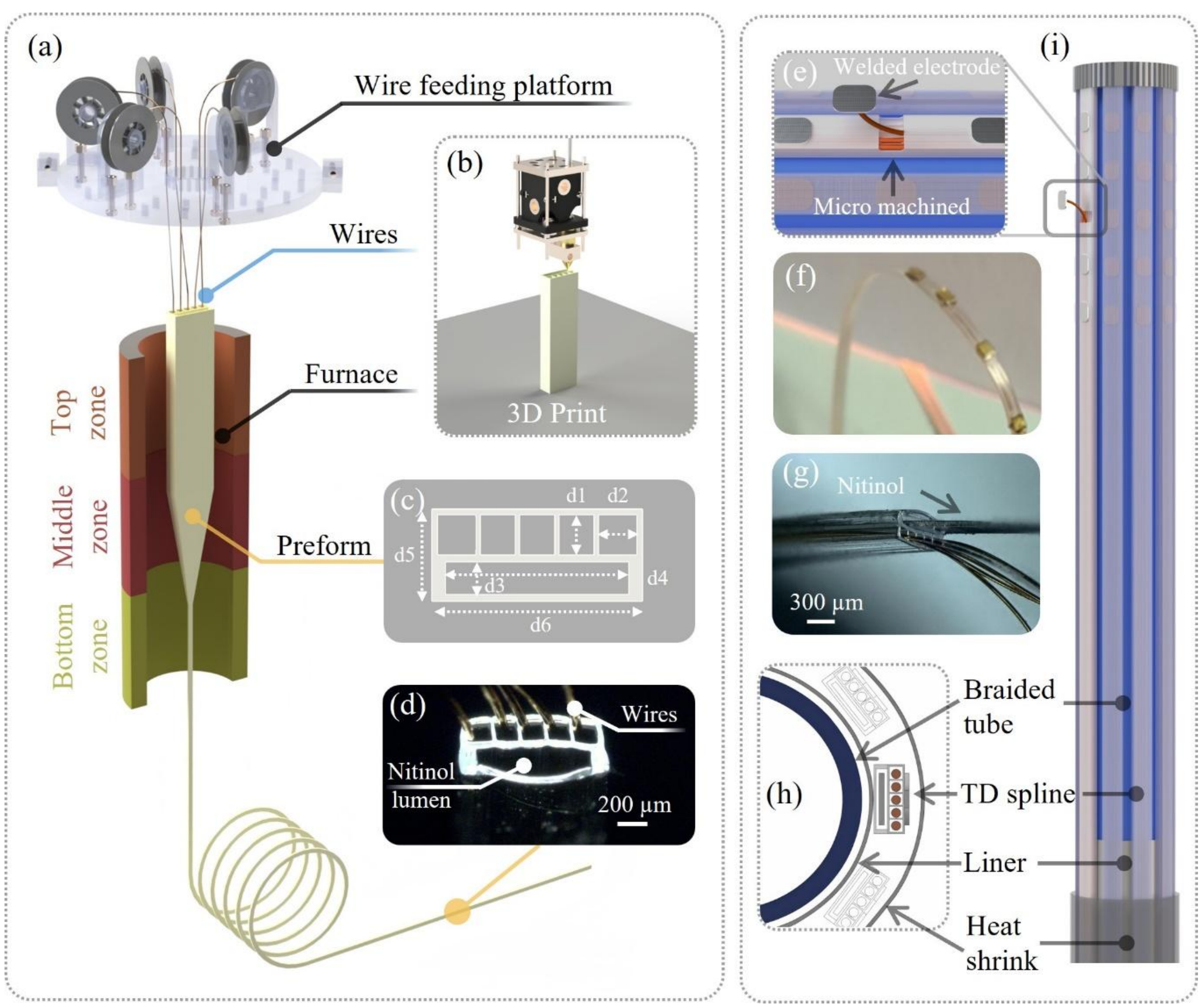

**Figure 2. Spline fabrication and mapping sheath assembly.** (a) Schematic illustration of the fiber drawing process of the spline with wire feeding. (b) 3D printing of the preform with designated cross-section prepared for spline fabrication. c. Cross-section and dimensions of the preform design, where d1=5.9, d2=5.9, d3=5.1, d4=28, d5=14, and d6=32mm. (d) Labeled microscopic cross-section view of the drawn spline with five conductive wires embedded. (e) Schematic representation of electrode integration in spline assembly. (f) Microscopic view of the spline with electrodes assembled. (g) Microscopic view of nitinol integration. (h) Cross-section illustration of the mapping sheath assembly, where TD spline represents a thermally drawn spline. (i) CAD representation of the mapping sheath distal end assembly.

### 2.3. Mapping Sheath Steering and Basket Deployment

By implementing the thermally drawn mapping splines and mapping sheath assembly described above, the system is integrated with a bespoke handle that facilitates the manipulation in an operating room (Figure 3a). The handle is customized and fabricated with 3D printing to organize the wire routing and add an axial driving mechanism for basket deployment (Supplementary Figure S1). The wires exposed from mapping splines are bundled and individually soldered to a connection cable with pins for plugging into commercial signal reading systems. The port at the handle's proximal end facilitates exposure of the connection cable and passage of an ablation catheter therein. The full assembly of the mapping sheath system's proximal end for manual manipulations is shown in Figure 3b.

With the handle's intuitive design, steering and basket deployment at the distal end of the sheath can be performed with a single hand within a small area on the operating bed (Figure 3c). With a rack-and-pinion mechanism, our design allows basket deployment and retraction with a single knob, owing to its minimal axial displacement requirement for spline deployment. Steering of the distal end of the mapping sheath is assisted by the ablation catheter, which is introduced into its central channel. By rotating the knob on the deflectable ablation catheter, the tip of the ablation catheter is steered, driving the distal segment of the mapping sheath that has a softer bending stiffness. The positioning of the ablation catheter can further control the length of the steering segment, hence offering a more agile manipulation method for surgeons.

The deployment of the basket structure enables versatile configurations, allowing more precise control of the basket shape to accommodate complex intracardiac contours. The deployment mechanism enables the basket diameter to expand from 3.7 mm in the undeployed state to 28 mm when fully deployed, comparable to the 3–22 mm range of a commercial comparator device with a mapping basket (*28*). Benefiting from the knob control mechanism, the deployment process allows adjustment of the deployment amplitude, as shown in Figure 3d. In the standard procedure hypothesized for clinical applications, the mapping sheath is introduced into the right atrium, steered to the target region, and the basket is deployed for mapping. As demonstrated in Figure 3e, this procedure can be achieved through cooperation between the designated steering and deployment mechanisms. The detailed deployment procedure is shown in Supplementary Video S1.

The steering performance of the mapping sheath, as shown in Figure 3f, was further characterized through workspace measurements. The results indicate that, with a 90 mm steering section, the sheath achieves a one-directional steering range of ±55 mm (Figure 3g). As shown in Figure 3h, the reconstructed three-dimensional workspace demonstrates that a hemispherical region with a radius of 55 mm can be reached through actuation of the handle knob in combination with rotation of the mapping sheath shaft. Detailed characterization procedures are provided in Supplementary Figures S2 and S3. The detailed steering procedure is shown in Supplementary Video S2.

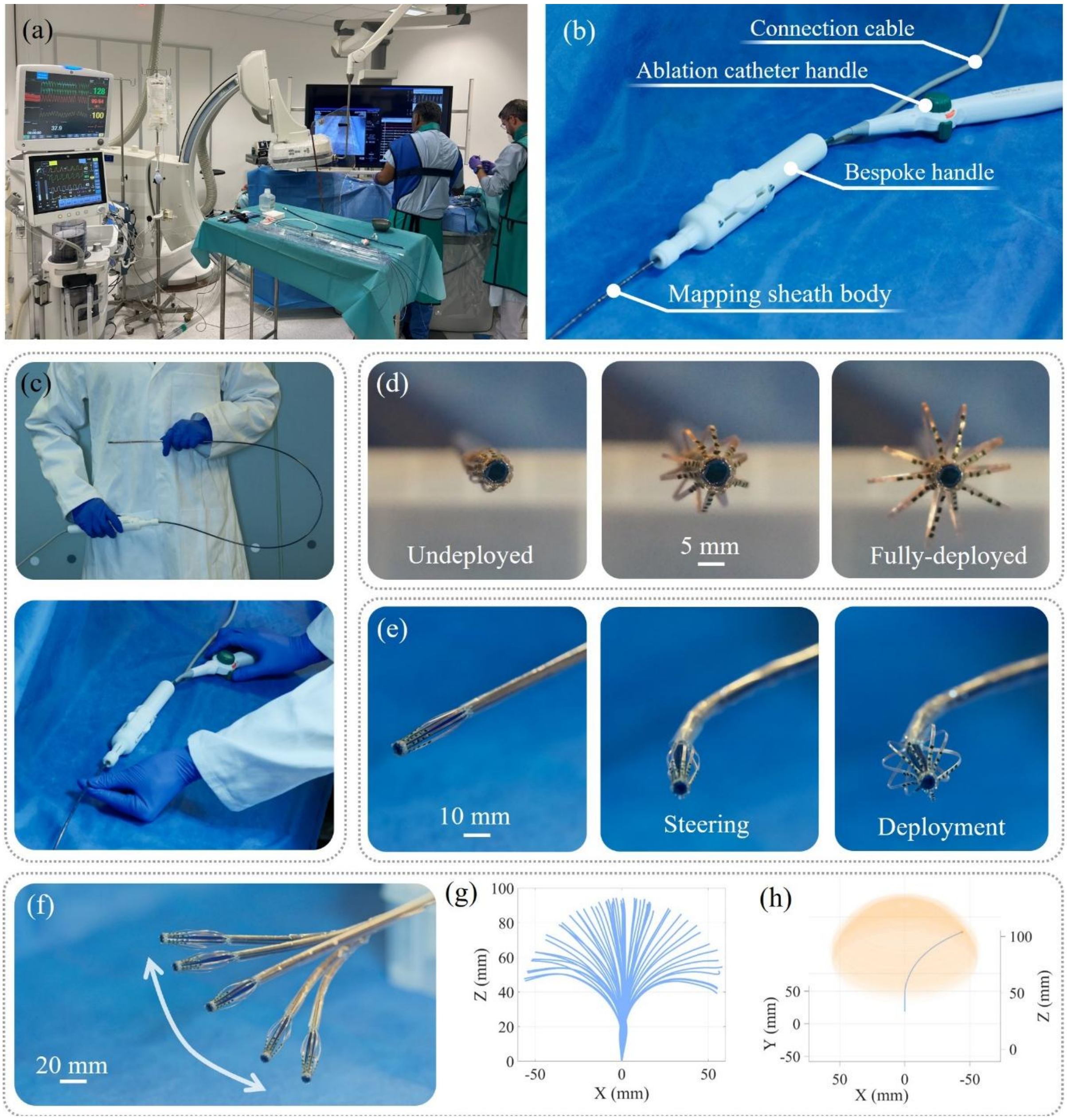

**Figure 3. Cardiac electrophysiology and detail views of mapping sheath functionalization.** (a) Photograph of a cardiac electrophysiological study on a porcine model within the cardiac lab suite. (b) Handle assembly, ablation catheter integration, and positioning of ablation catheter steering handle. (c) Manual manipulation of mapping sheath's basic functions: steering and deploying. (d) Front view of the basket deployment under various statuses and the cooperation with the steering mechanism. (e) Isometric views of the mapping sheath distal end showcasing a standard steer and deploy procedure. (f) Isometric view of the mapping sheath steering. (g) Workspace measurements of the mapping sheath in 2D. (h) Workspace measurements of the mapping sheath in 3D.

## 2.4. Bench-top Mechanical Characteristics for Mapping Sheath Assembly

A rigorous benchmarking of the mapping sheath assembly was conducted through a comprehensive experimental assessment of its mechanical performance, encompassing flexibility (flexural rigidity), pushability (axial rigidity), and deployability. The samples tested in the bench-top experiments are presented in Supplementary Table S2. The flexibility is defined with the equation:

$$bending\ stiffness = \frac{radial\ force}{flexural\ displacement}.$$

The pushability and deployability are defined with the equation:

$$pushability = \frac{axial\ force}{axial\ displacement}.$$

Flexibility quantifies the bending stiffness of the endovascular devices and is a key determinant of their ability to guide and support intraluminal interventions while maintaining distal compliance. An experimental setup was built as illustrated in Supplementary Figure S4. Figure 4a presents the bending stiffness measurements of the mapping sheath shaft compared with those of a commercial mapping catheter (*29*). Given the equation described above, the shaft body of the mapping sheath exhibits an approximate 35 % lower stiffness when compared with the commercial sample, calculated as 388 ± 12 mN/mm and 526 ± 24 mN/mm. The distal section of the mapping sheath demonstrates a similar stiffness compared with the commercial one, calculated as 129 ± 7 mN/mm and 134 ± 10 mN/mm. With our dual-stiffness design of the braided tube, the mapping sheath assembly achieves a comparable bending stiffness at the distal end for steering while retaining high stiffness in the body section for endovascular navigation. Detailed measurements are illustrated in Supplementary Figure S5.

Figure 4b illustrates the flexibility characterization of splines. The full mapping spline assembly calculated with a bending stiffness of 0.2 ± 0.002 mN/mm demonstrates a 75.3% higher stiffness than the drawn spline (0.110 ± 0.002 mN/mm) and a 30.5% lower stiffness than the spline of a commercial basket catheter (*28*) (0.290 ± 0.002 mN/mm). Supplementary Figure S6 and Figure S7 showcase the procedure of the experiment.

Pushability evaluates the spline's ability to deploy and apply force to tissues, which is crucial for quantifying the axial force required for deployment while avoiding tissue damage. An experimental setup was built as illustrated in Supplementary Figure S8. The axial forces increase dramatically during the initial axial displacement (< 20% of the spline length) and then stabilize, consistent with the expected behavior in which axial compression rises rapidly until the structure reaches its Euler buckling threshold, after which deformation proceeds in a stable post-buckling regime requiring an approximately constant force (Figure 4c). The full deployment of a single mapping spline (201 mN/mm pushability) requires a 44.6 % higher axial force and a 49.8% lower axial force compared with the thermally drawn spline (139 mN/mm pushability) and commercial basket spline (301 mN/mm pushability), respectively. The results show that our bespoke mapping spline showcases a higher pushability, allowing more spline integration with the same axial force required.

Deployability defines the ease of deploying the basket structure at the mapping sheath's distal end. Figure 4d characterizes the axial deployment force applied with various statuses of single spline and basket deployment. Due to the early appearance of Euler buckling threshold, half and full spline deployments require a similar axial force around 201 ± 5 mN. The full basket assembly with ten mapping splines integrated shows a minimum force of 1930 ± 25 mN exerted from the co-axial sliding of sheath bodies in full deployment. Detailed sequential images of the characterization are shown in Supplementary Figure S9.

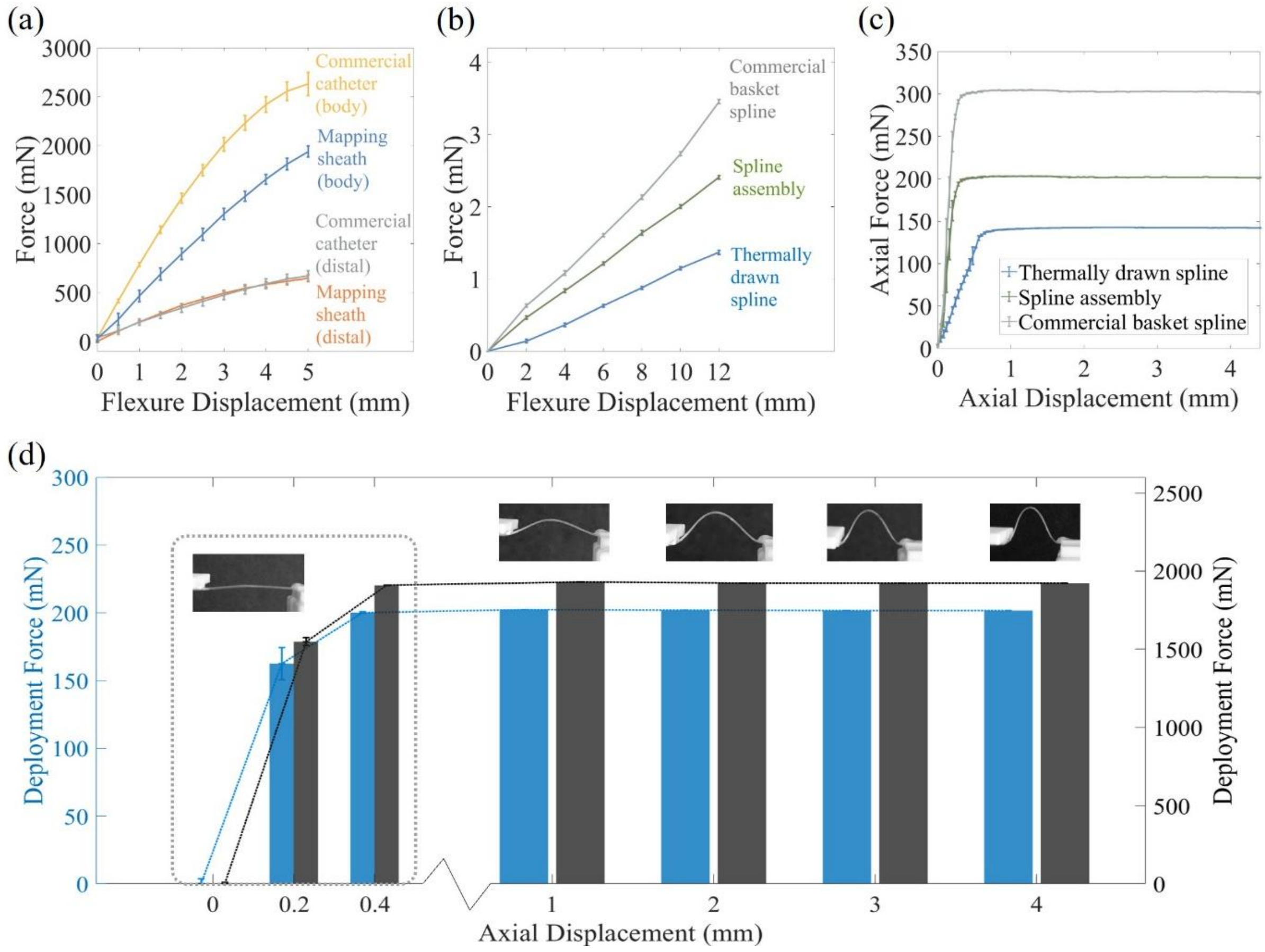

**Figure 4**. **Benchmark mechanical evaluation results for the mapping sheath assembly.** (a) Flexural rigidity test on the main body and the distal section of the commercial mapping catheter and our bespoke mapping sheath. (b) Flexural rigidity test on single spline, performed on thermally drawn spline, spline assembly, and commercial basket spline (INTELLAMAP ORION$^{TM}$, Boston Scientific). (c) Pushability test on the thermally drawn spline and the spline assembly. (d) Axial deployment force characterization on single spline (left y axis) and full basket assembly (right y axis) with certain steps, where the blue axis represents the single mapping spline and the black axis represents the basket assembly.

### 2.5. Bench-top Electrophysiological Characteristics for Mapping Splines

To evaluate the signal recording performance of our mapping splines, we conducted electrophysiological experiments on a bespoke characterization setup. A series of electric currents with morphology and amplitudes programmed by an external waveform generator was applied to a tissue-like agar, simulating intracardiac wavefront propagation (Figure 5a). Figure 5b illustrates our electrode output in response to triangle waveforms with peak amplitudes from 3 mV to 10 mV measured by a benchmark oscilloscope. The waveform amplitudes match the voltage variances in intracardiac electrograms (*30*). The sensitivity of the electrodes is defined as: *Sensitivity = Sensing Amplitude/Programmed Amplitude*. In comparison to the measurement from a commercial mapping catheter (*31*) with a sensitivity of 0.63, our bespoke electrode spline exhibits a comparable sensitivity of 0.49 despite having a 73 % smaller contact sensing area (Figure 5c). Figure 5d further shows a sensitivity variance of ± 0.05, across 50 electrodes, which follows a Gaussian distribution with $\mu = 0.49$ and $\sigma = 0.04$. In typical voltage range of intracardiac signals (2-10 mV), our mapping splines exhibit a signal-to-noise ratio of 8.3 dB to 21.7 dB and a correlation coefficient over 0.95 by comparing the programmed waveforms (Figure 5e).

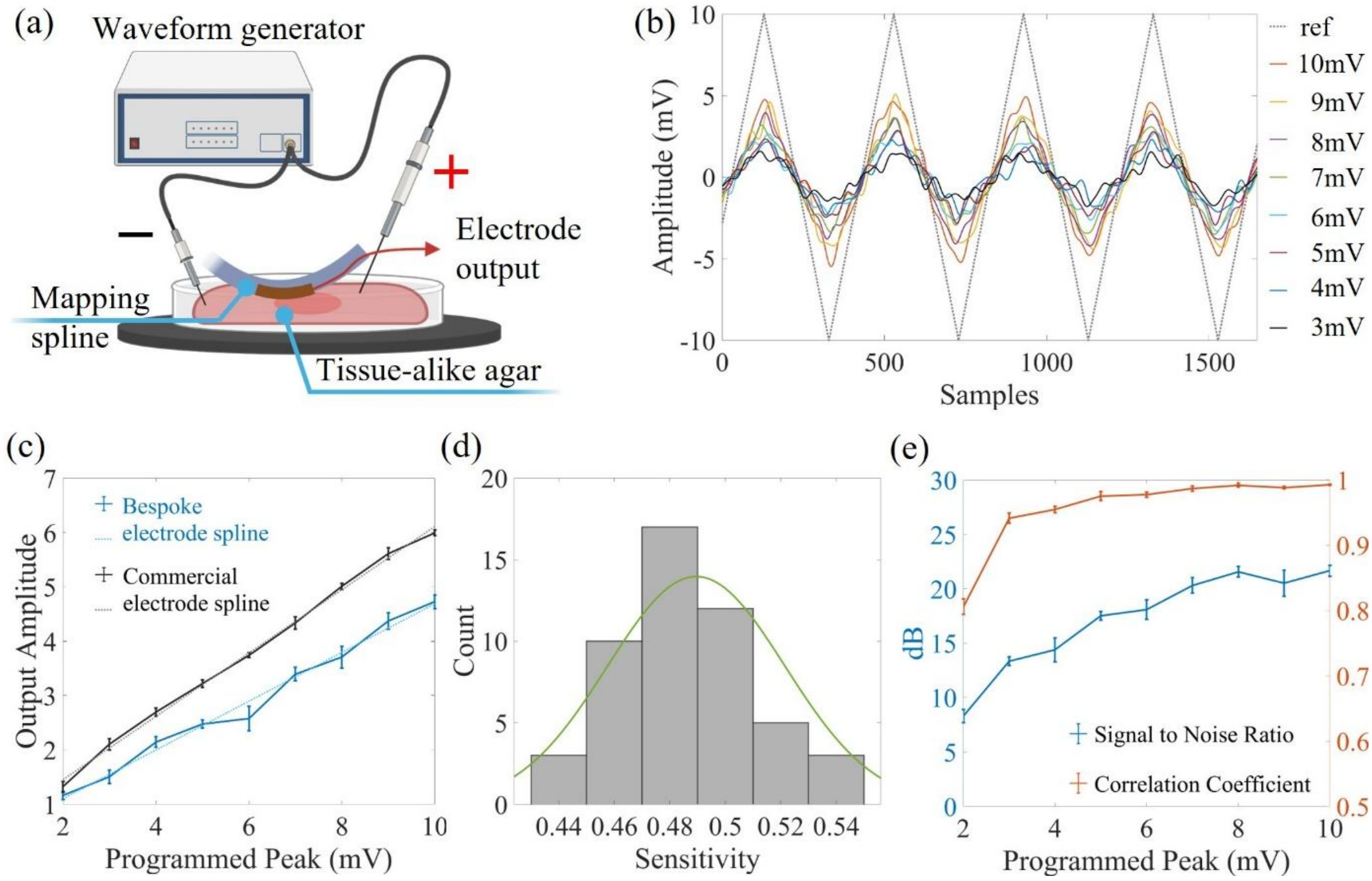

**Figure 5. Mapping spline characterizations.** (a) Schematics of the characterization setup. (b) Amplitudes of the signal recording under different applied voltages. (c) Output voltage responses to signals with various amplitude generated by the waveform generator. (d) Sensitivity characterizations and the Gaussian approximation with a batch of tested electrodes. (e) Signal to noise ratio and correlation coefficient in comparison to reference signals.

### 2.6. In-vitro Phantom Study for Mapping Sheath

The clinical applicability of the mapping sheath was assessed in various surgical scenarios with an in vitro phantom study. The phantom experiments were conducted using a normal adult left atrium (LA) phantom, resembling the catheter pathway during cardiac interventions (Figure 6a). The details of the phantom study procedures are illustrated in Supplementary Video S3. As shown in Figure 6b, the mapping sheath was introduced via the inferior vena cava and advanced to a total length of 33 cm, after which a commercial ablation catheter was inserted into its central channel. With the tip maneuverings driven by the steerable ablation catheter, the distal end of the mapping sheath gained access to LA through a single opening, representing a single transseptal puncture of the interatrial septum.

The maneuverability was further confirmed by guiding the mapping sheath to reach the anatomies that are usually critical in mainlining arrhythmias, including the right pulmonary veins (RPV), left pulmonary veins (LPV), and posterior wall. Figure 6c demonstrates that the mapping sheath can be steered into the veins in both RPV and LPV. The mapping splines were then successfully deployed within these areas, bringing the electrodes into conformal contact with the intracardiac wall. The mapping sheath was also successfully advanced to and deployed along the posterior wall of the LA. This demonstrates that our bespoke mapping sheath can probe various anatomies within the LA, integrate ablation capability, and enable agile deployment within the heart.

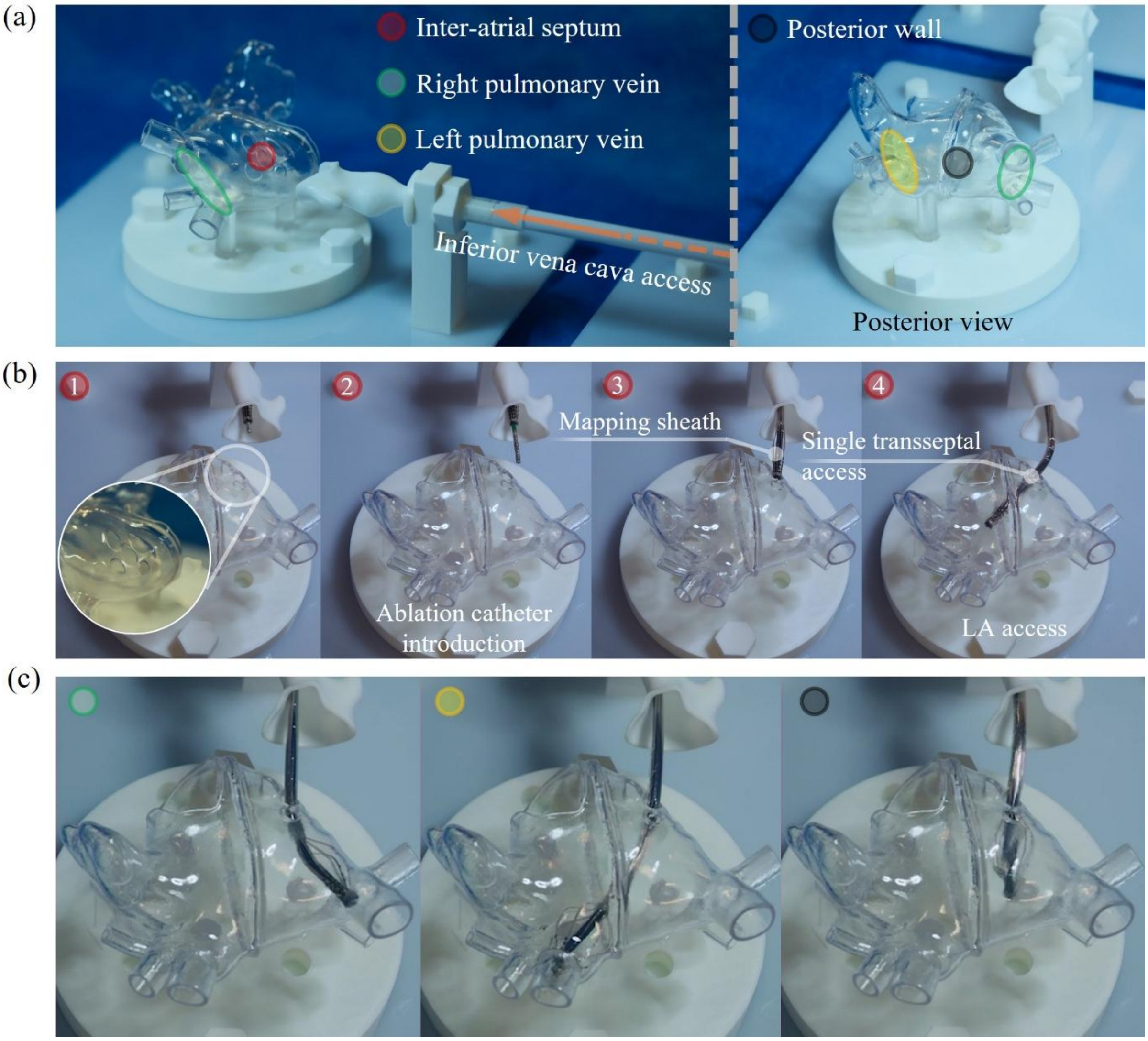

**Figure 6. In vitro phantom study.** (a) Left atrium (LA) phantom with left pulmonary veins (LPV), right pulmonary veins (RPV), inter-atrium septum, and inferior vena cava. (b) Image sequence demonstrating the navigation of the mapping sheath and introduction of the ablation catheter into the LA with access from the inferior vena cava. (c) Images

illustrating the mapping sheath probing with the mapping basket deployed at the RPV (green), LPV (yellow), and posterior wall (black).

**2.7. Ex vivo studies with a Langendorff porcine heart model.**

An ex vivo electrophysiological study was conducted in a Langendorff-perfused porcine heart model to evaluate the device's capability for cardiac electrogram acquisition. Figure 7a shows the schematic of the Langendorff apparatus, with details of the setup provided in Supplementary Figure S10. Before the experiment, a porcine heart was retrogradely perfused through the aorta with physiological solution, allowing it to continue beating in an ex vivo condition (*32*). A pacing electrode was attached to pace the heart at a cycle length of 660 ms, corresponding to approximately 90 beats per minute. A commercial coronary sinus (CS) mapping catheter was positioned within the CS. To maintain contact between the mapping electrodes and the epicardium, the mapping spline was gently pressed against the epicardial surface using a soft foam pusher (Figure 7b). In addition, a commercial EP catheter (*29*) was placed adjacent to the testing spline and used as a reference for signal quality comparison.

Corresponding epicardial electrograms were recorded during the electrophysiological study. As shown in Figure 7c, unipolar electrograms recorded during sinus rhythm using the bespoke mapping spline exhibited highly comparable signal morphology to those obtained from the electrode of the commercial mapping catheter (*31*). A high correlation coefficient of 0.99 between the two recordings was calculated after compensating for a 7-sample temporal delay, which statistically shows the similarity in signal morphology. The signal-to-noise ratios (SNRs) derived from the recorded signals were 26.6 dB and 26.7 dB for the bespoke mapping spline and the commercial EP catheter, respectively, indicating comparable unipolar signal fidelity.

Bipolar electrograms were further reconstructed by calculating the differential signals between adjacent electrode pairs (Figure 7d). The bespoke mapping spline successfully resolved local cardiac activation potentials, as evidenced by the sharp activation deflections in the electrograms, with sharp electrogram deflections temporally aligned with the CS catheter and adjacent commercial EP catheter. Compared with the unipolar recordings, the bipolar electrograms demonstrated improved SNRs of 31.7 dB and 33.5 dB for the bespoke mapping spline and commercial catheter, respectively. This improvement is attributable to the suppression of far-field components and common-mode noise through bipolar signal derivation. Overall, the ex vivo electrophysiological results demonstrate that the bespoke mapping spline provides sufficient electrogram signal quality, both in terms of noise performance and local activation delineation, for cardiac electrogram recording.

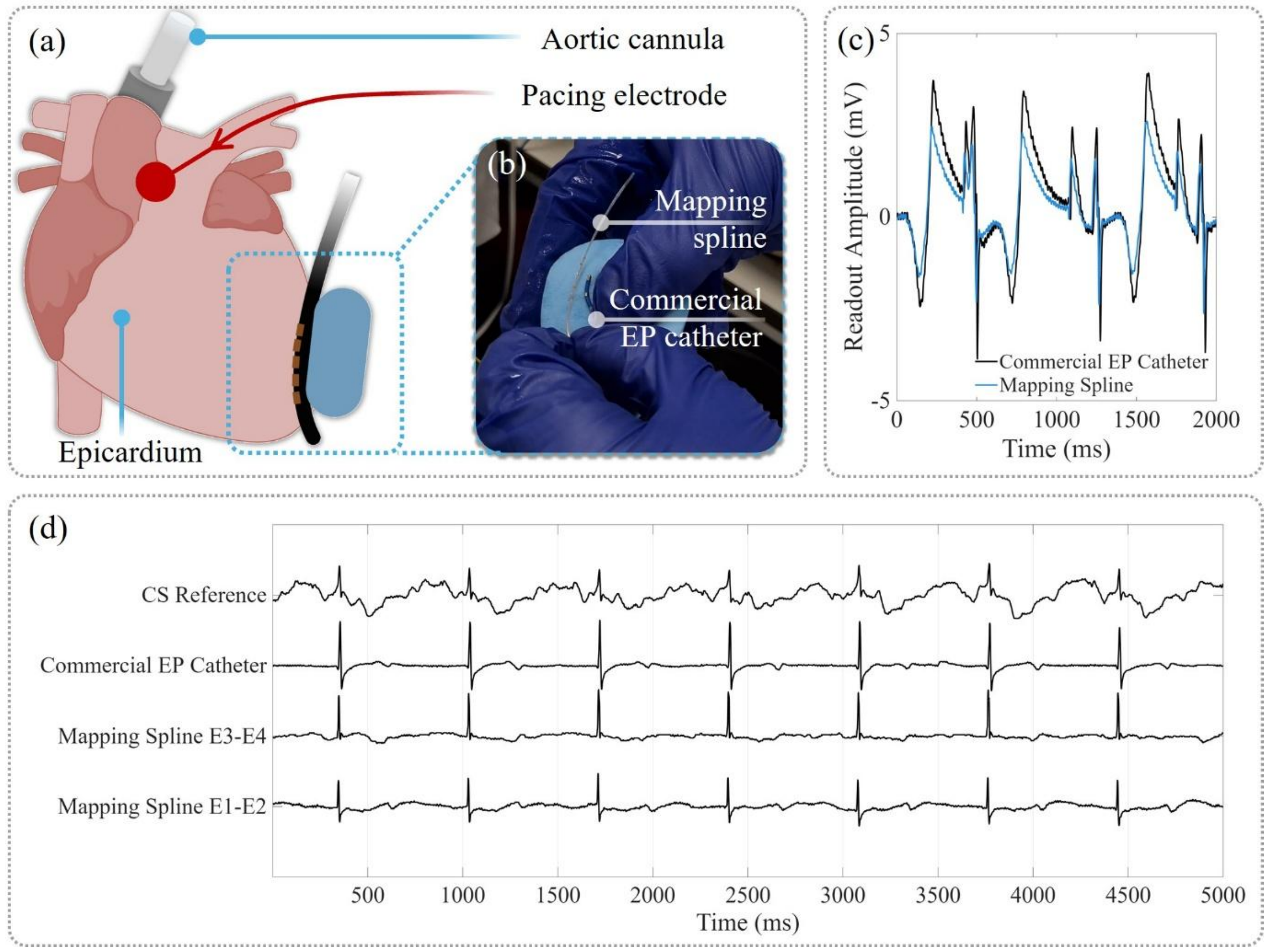

**Figure 7. Ex vivo epicardial electrophysiological studies on Langendorff porcine heart model.** (a) Schematic of the experimental setup for maintaining nutrient supply via an aortic cannula controlling heat rhythm via a pacing electrode. (b) Image showcasing the catheter placement. (c) Unipolar electrogram readouts from commercial EP catheter and bespoke mapping spline. (d) Real-time bipolar electrogram readouts from CS reference, commercial EP catheter and bespoke mapping spline.

### 2.8. In vivo studies of the mapping sheath with a porcine model.

Following ex vivo electrogram recording tests in the Langendorff porcine heart model, we conducted animal in vivo porcine experiments to further validate the electrophysiological and mechanical functions of the mapping sheath under (Figure 8a). The mapping sheath was encapsulated within a 14 Fr sheath and introduced through a 16 Fr introducer, which provided vascular access and prevented blood backflow at the percutaneous access site (Figure 8b). The mapping sheath cables, which transmitted electrode signals, were connected to a commercial cardiac EP recording system (*33*), as shown in Supplementary Figure S11. Detailed sequential fluoroscopic images of the in vivo procedure are provided in Supplementary Video S4.

During the in vivo experiment, the pig was endotracheally intubated, placed under general anesthesia, and positioned supine. The procedure was initiated by inserting the mapping sheath into the 14 Fr sheath, which provides sufficient rigidity for catheter introduction. The 14 Fr sheath was then inserted through a 16 Fr introducer and advanced to the entrance of the inferior vena cava. The mapping sheath was subsequently advanced until its distal end and the splines were fully exposed, as shown in the fluoroscopic image in Figure 8c. A steerable decapolar catheter (Deca) was introduced into the central lumen of the mapping sheath and used to guide the sheath into the RA (Figure 8d). A magnified fluoroscopic view of this procedure is shown in Figure 8e. Once the mapping sheath reached the RA, the liner was retracted to pull back the distal end and drive outward bending of the splines, thereby forming the mapping basket within the heart (Figure 8f). The mapping sheath was able to be deflected in both deployed and undeployed states via steering the Deca therein. As shown in Figure 8g, the mapping sheath was steered to the superior aspect of the RA, where the mapping splines were deployed to conform to the endocardial wall and facilitate stable electrode contact for electrogram recording.

Corresponding bipolar intracardiac electrograms were recorded during this procedure. As shown in Figure 8h, the recorded signals exhibited clearly identifiable activation patterns, demonstrating the capability of the mapping sheath to acquire endocardial electrograms suitable for cardiac electrophysiological analysis. Figure 8i presents the amplitude map for each electrode pair, derived from the mean peak amplitude across three consecutive activation cycles. Electrode pairs located closer to the distal end of the mapping sheath exhibited larger signal amplitudes, with the two distal electrode pairs showing an approximately 52% greater mean amplitude (~0.38 mV) than the two proximal electrode pairs (~0.25 mV). Using a 0.1 mV threshold as an indicator of effective tissue contact, the two proximal electrode pairs on adjacent splines, splines 1 and 5, showed limited electrogram acquisition. A local activation map was then reconstructed from these electrograms (Figure 8j), revealing activation propagation from the distal toward the proximal end of the mapping sheath. These findings demonstrate that the mapping sheath enables localized electrophysiological mapping and can resolve the direction of wavefront propagation.

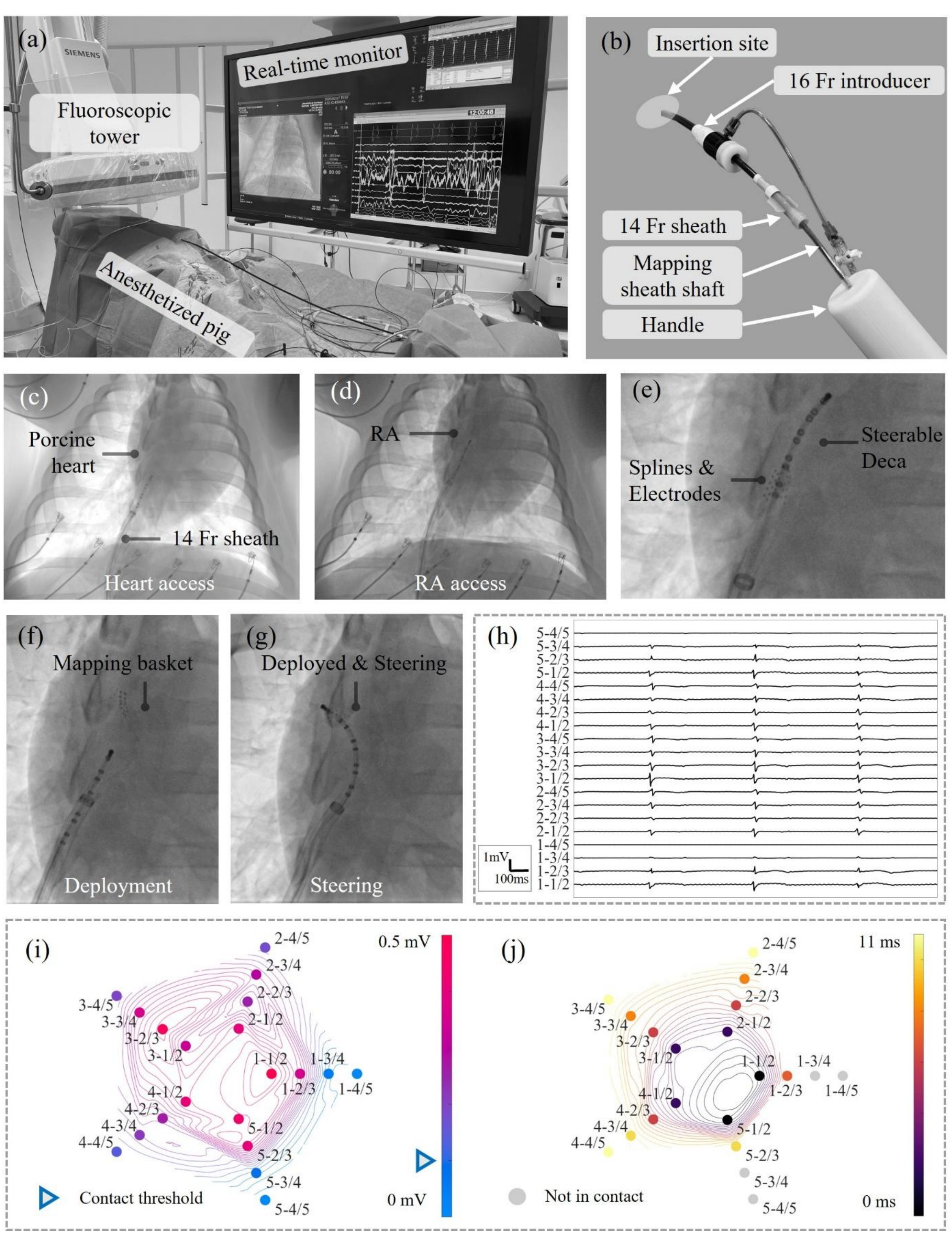

**Figure 8. In vivo demonstrations and electrophysiological evaluation of the mapping sheath in porcine model.** (a) Overview of the experimental setup. An endotracheally intubated pig was maintained under general anesthesia in the supine position. A fluoroscopy system provided real-time X-ray imaging of the thoracic cavity during the procedure, while a monitor displayed live fluoroscopic images and intracardiac

electrogram recordings. (b) Catheter introduction setup showing vascular access through the right femoral vein. (c) Fluoroscopic images of the mapping sheath being advanced toward the heart. (d) Fluoroscopic view of the mapping sheath that has gained access to the right atrium. (e) Magnified fluoroscopic view showing the steerable decapolar catheter and the electrodes of the mapping sheath, visible as radiopaque dot arrays. (f) Fluoroscopic view showing deployment of the mapping basket. (g) Fluoroscopic view showing deployment and steering of the mapping sheath for tissue contact. (h) Representative bipolar intracardiac electrogram recordings acquired during the procedure. (i) Voltage amplitude map reconstructed from the collected electrograms. (j) Local activation map reconstructed from the collected electrograms.

## Discussion and Conclusion

An unmet need in interventional electrophysiology is a compact device that can combine adaptive multielectrode mapping with immediate ablation catheter delivery through a single access route. Current procedures often require separate mapping and ablation catheters, repeated device exchanges, or dual transseptal access, which can increase procedural complexity and delay treatment of dynamic arrhythmogenic substrates. In this work, we developed a deployable mapping sheath that integrates a circumferential multipolar electrode basket with a central lumen compatible with standard ablation catheters, providing a unified platform for local cardiac mapping and therapeutic catheter delivery.

Thermal fiber drawing enabled the fabrication of ultra-thin mapping splines with embedded conductors and integrated mechanical reinforcement. The resulting splines supported closely spaced PtIr electrodes, maintained structural integrity during deployment, and were assembled into a low-profile basket capable of controlled expansion. At the device level, the mapping sheath preserved an 8 Fr working lumen within a 14 Fr introducer compatible profile, allowing integration with steerable ablation catheters. Its dual-stiffness shaft provided proximal support for navigation and distal compliance for steering within a range of ±55 mm, while the basket could be deployed over various diameters (3.7-28 mm) and retracted with a bespoke handle. Together, these results demonstrate that thermal fiber drawing provides a rapid, scalable, and cost-effective platform for prototyping miniaturized, mechanically compliant intravascular devices with complex architectures and integrated functionality.

Mechanical and electrophysiological validation confirmed that the mapping sheath provides the structural performance and recording capability required for intracardiac use. The shaft and splines exhibited mechanical properties compatible with catheter navigation, steering, and controlled basket deployment. In an in vitro LA phantom study, the sheath was advanced along an inferior vena cava to LA pathway through a simulated interatrial septal access site and was deployed at key anatomical regions, including the pulmonary veins and posterior wall. These findings support the procedural feasibility of the device for intracardiac mapping. In terms of the recording capability, the electrodes reproduced programmed waveforms across voltage ranges representative of intracardiac electrograms, with consistent sensitivity across electrodes. In an ex vivo Langendorff-perfused porcine heart model, unipolar and bipolar recordings from the bespoke mapping splines showed clear local activation deflections and favorable signal-to-noise performance. These findings support the feasibility of using thermally drawn mapping splines for reliable electrogram acquisition from cardiac tissue.

An in vivo porcine study further demonstrated vascular introduction, fluoroscopic visualization, intracardiac deployment, steering, tissue contact, and acquisition of electrograms within the heart. Reconstructed voltage and activation maps further showed that the sheath could identify regional differences in electrode contact and resolve local activation propagation, indicating that the acquired signals were sufficient not only for electrogram recording but also for spatial electrophysiological interpretation. Together, these findings support the translational potential of the device for future applications in ablation catheter delivery and intracardiac mapping.

Despite the advantages of this mapping sheath design, the device can be further optimized and validated before translation into human clinical studies. Increasing electrode density may improve spatial resolution in intracardiac mapping and enhance delineation of complex activation patterns. Further clinical validation can focus on assessing coordinated mapping and ablation performance, as well as the device's ability to monitor surrounding activation patterns during pacing. In addition, although the device demonstrated feasibility in a controlled preclinical animal model, anatomical, physiological, and procedural differences between animal models and patients must be carefully addressed to ensure reliable performance in clinical use.

Overall, this work highlights the clinical potential of a novel mapping sheath integrating electrophysiological recording and ablation catheter delivery. The thermal drawing technology provides a valuable and flexible tool for medical device fabrication, offering advantages in customization, miniaturization, rapid prototyping, cost-effectiveness, and manufacturing scalability. The unique mapping sheath design holds great potential, empowering cardiac electrophysiologists with a new approach to deliver diagnosis and treatment on an integrated, compact platform.

## Methods

**Preform fabrication.** Spline preforms were designed by first using computer-aided design software (SolidWorks 2020, Dassault Systèmes, France). The optimized preform was 170 mm long and had a cross-sectional profile scaled 40-fold relative to the final spline geometry. The preforms were printed from polycarbonate filaments using a fused deposition modeling printer (Ultimaker 3 Extended, Netherlands). The printed preforms were then consolidated in a vacuum oven at 70°C to remove residual moisture and reinforce the interlayer structure. Detailed printing parameters are listed in Table 1 of (*34*). Before fiber drawing, enameled copper wires (60 µm in diameter) mounted on a wire feeding platform were inserted into each channel of the preform.

**Fiber drawing.** Fiber drawing was performed using a custom-made fiber drawing tower comprising a three-zone heating furnace, a preform feeding platform, a wire feeding platform, and a fiber-pulling capstan (Supplementary Figure S12). The process was initiated by securing the preform in a polyether ether ketone (PEEK) reinforced holder (Bay Plastics Ltd., UK) with a diameter of 40 mm. Because the preforms were fabricated layer by layer and therefore had lower tensile strength than molded or extruded solid preforms, a lower drawing temperature was used during drawing than is typically applied to solid preforms (*22*). Initial draw temperatures of 120°C, 230°C, and 85°C were applied to the top, middle, and bottom zones, respectively, to reduce the preform viscosity by several orders of magnitude. A neck-down region was induced under a constant tension of 15g, together with the weight of the lower end of the preform. Once the lower end of the preform exited the furnace, the thickened distal portion was removed. Afterward, the resulting fiber was connected to the fiber-pulling capstan, and the draw temperature was gradually reduced to 195°C. The preform feeding speed then remained at 2mm/min. By controlling the capstan rotation speed and monitoring the outer diameter of the drawn fiber using a laser sensor, the fibre pulling speed then remained at 1.55m/min and fibers with a consistent thickness were obtained.

**Electrode fabrication.** PtIr electrodes were fabricated by importing the two-dimensional CAD design into the KYLA laser control software (Liège, Belgium) to define the laser-machining path. The pattern consisted of 20 inside-out linear contour layers with a 4-µm pitch along the electrode perimeter and a clockwise writing direction. A 100-µm-thick PtIr foil (70% Pt, 30% Ir; Merck KGaA, Germany) was affixed to a 5-mm-thick stainless-steel plate using double-sided tape and cut along the predefined path using an OPTEC femtosecond laser system with a 515-nm wavelength (Liège, Belgium). The laser parameters were a 400-fs pulse width, a 50,000-Hz repetition rate, 120- and 280-μs laser on/off delays, respectively, and 40% output power. After laser cutting, individual electrodes were released from the foil and cleaned with isopropanol to remove carbonized debris and adhesive residue.

**Electrode spline assembly.** Spline samples were selected from the drawn fibers based on the following criteria: (i) thickness within a ±30μm tolerance, (ii) preservation of cross-sectional architecture, and (iii) intact enameled wires. The top surface of each spline, corresponding to the thin wall adjacent to the wire channels, was then removed according to the intended electrode layout. The wires were subsequently exposed and laser-welded to the PtIr electrodes. Electrode assembly was completed by sealing the electrodes to the spline through microinjection of a medical-grade ultraviolet-curable adhesive (Loctite AA3321, Henkel AG & Co. KGaA, Germany). In the final assembly step, a 50-mm-long flat Nitinol ribbon (600 μm wide and 100 μm thick) was inserted into the Nitinol lumen to form the distal end of the deployment mechanism.

**Mapping sheath assembly.** The mapping sheath was assembled by coaxially sliding an etched PTFE liner tube (3.45 mm ID, 3.55 mm OD) over a variable-durometer PEBAX braided tube (2.85 mm ID, 3.25 mm OD) until the liner was positioned 30 mm proximally to the distal end of the braided tube. One assembled electrode spline was then bonded to the distal end of the braided tube using ultraviolet-curable adhesive. The remaining nine electrode splines were bonded in the same manner, with equal circumferential spacing around the braided tube. At a position 35 mm proximal to the distal spline-bonding site, the splines were further bonded to the PTFE liner. This differential fixation of the splines to the braided tube and liner formed the basket-deployment mechanism, enabling spline expansion through relative axial displacement between the sheath layers. A thin-walled PET heat-shrink tube (5 mm OD, 3.8 mm recovered diameter, 0.015 mm wall thickness) then slid over the outer surface of the sheath to fully cover the liner. After heating at 120°C until complete recovery of the heat-shrink tube, the sheath shaft was assembled. The mapping sheath assembly was completed by attaching the sheath shaft to a bespoke handle, as described in Supplementary Figure S1.

**Phantom study.** The phantom was a 3D printed left atrium with a simulated inferior vena cava (from Hongchuang Jingwei Technology Co., Ltd). The phantom study was performed by inserting the mapping sheath into the inferior vena cava entrance of the left atrial phantom. After the sheath was advanced through the inferior vena cava pathway, an 8Fr steerable catheter was inserted into the working channel of the sheath. The sheath was then guided through one of the openings in the simulated interatrial septum by proximal shaft rotation and tip steering controlled by the handle knob of the steerable catheter. Using the same maneuvering approach, the distal end of the mapping sheath was sequentially steered toward predefined anatomical targets, including the right and left pulmonary veins and the posterior wall. The mapping splines were deployed by rotating the knob on the bespoke handle, which controlled the relative axial displacement between the braided tube and the liner, thereby inducing spline bending. After completion of the phantom study, the mapping sheath was withdrawn from the phantom by pulling the proximal end.

**Langendorff experiments.** Signal quality of the developed electrode spline was evaluated in an ex vivo porcine beating heart preparation adapted from a Langendorff-perfused whole-heart model (*32*). An intact heart from healthy large white pigs (70 to 80 kg, 4 to 5 months old) was explanted, transported on ice in cold cardioplegic solution, and mounted within 1 hour on a custom Langendorff perfusion system. The heart was reperfused by retrograde aortic perfusion with oxygenated Tyrode's solution in a 5-liter recirculating circuit incorporating a heating coil, bubble trap, and high-flow peristaltic pump (Supplementary Figure S13). Perfusate temperature was maintained at 37 ± 0.5°C. After washout of cardioplegia and a stabilization period of 15 to 30 min, the heart recovered organized electrical and mechanical activity suitable for electrophysiological assessment. Throughout the experiment, electrocardiogram, pressure, and temperature were continuously monitored. The investigational electrode spline was positioned on the epicardial surface and maintained contact with the beating myocardium using a form pusher to ensure stable, reproducible tissue contact. A commercial mapping catheter (AFocus$^{TM}$) placed adjacent to the test site served as the primary reference. In addition, a coronary sinus mapping catheter was introduced to provide an independent intracardiac reference signal. Simultaneous recordings from the spline and reference catheters were used to compare electrogram quality under identical perfusion conditions.

**Animal experiments.** One adult pig was prepared for orotracheal intubation in accordance with Directive 2010/63/EU. The animal experiment was performed at IHU Liryc France. The animal was anesthetized, endotracheally intubated, and maintained under mechanical ventilation with inhaled isoflurane and intravenous propofol. Electrocardiography, blood pressure, and oxygen saturation were continuously monitored throughout the procedure. Intracardiac electrograms were collected on a commercial electrophysiology recording system (LabSystem Pro, Boston Scientific). Vascular access was obtained by femoral vein cannulation under fluoroscopic guidance, followed by placement of a 16Fr introducer. Before introduction, an 8Fr radiofrequency ablation catheter (THERMOCOOL, Johnson & Johnson MedTech Inc.) was inserted into the working channel of the mapping sheath. The combined mapping sheath and ablation catheter assembly was then inserted into a 14Fr sheath and delivered through the 16Fr introducer. Once visualized under fluoroscopy, the mapping sheath and ablation catheter were gradually advanced into the right atrium. Navigation of the mapping sheath within the right atrium was performed by steering the ablation catheter to facilitate intracardiac tissue contact. The mapping splines were deployed within the heart to increase tissue contact. At the end of the experiment, the mapping sheath was withdrawn through the 16Fr introducer.